\documentclass[onecolumn]{elsart}
\usepackage [dvips] {graphicx}
\begin{document}

\begin{frontmatter}


\title{Detector Description and Performance\newline
for the First Coincidence Observations\newline between LIGO and
GEO}

\collaboration{The LIGO Scientific Collaboration\newline
{\scriptsize Corresponding Author: David Shoemaker, MIT NW17-161,
175 Albany St., Cambridge, MA 02139\newline Tel: 617 253 6411 ~~
Fax: 617 253 7014 ~~Email: {\tt dhs@ligo.mit.edu}}}

\author[CT]{B.~Abbott},
\author[LV]{R.~Abbott},
\author[LM]{R.~Adhikari},
\author[UW]{B.~Allen},
\author[FA]{R.~Amin},
\author[CT]{S.~B.~Anderson},
\author[TC]{W.~G.~Anderson},
\author[CT]{M.~Araya},
\author[CT]{H.~Armandula},
\author[CT]{F.~Asiri\thanksref{SLAC}},
% ...
\address[CT]{LIGO - California Institute of Technology, Pasadena, CA  91125, USA}
\address[LM]{LIGO - Massachusetts Institute of Technology, Cambridge, MA 02139, USA}
% ...
\address[LV]{LIGO Livingston Observatory, Livingston, LA  70754, USA}
% ...
\address[TC]{The University of Texas at Brownsville and Texas Southmost College, Brownsville, TX  78520, USA}
% ...
\address[FA]{University of Florida, Gainsville, FL  32611, USA}
% ...
\address[UW]{University of Wisconsin-Milwaukee, Milwaukee, WI  53201, USA}
% ...

\thanks[BALL]{Currently at Ball Aerospace Corporation}
\thanks[DEL]{Currently at University of Delaware}
\thanks[EC]{Currently at European Commission, DG Research, Brussels, Belgium}
\thanks[EGO]{Currently at European Gravitational Observatory}
\thanks[ESA]{Currently at ESA Science and Technology Center}
\thanks[HARV]{Currently at Harvard University}
\thanks[HOF]{Currently at Hofstra University}
\thanks[HPL]{Currently at HP Laboratories}
\thanks[IAP]{Currently at Institute of Advanced Physics, Baton Rouge, LA}
\thanks[INTC]{Currently at Intel Corp.}
\thanks[JPL]{Currently at NASA Jet Propulsion Laboratory}
\thanks[KECK]{Currently at Keck Observatory}
\thanks[LAPP]{Currently at Laboratoire d'Annecy-le-Vieux de Physique des Particules}
\thanks[LIGHTBIT]{Currently at LightBit Corporation}
\thanks[LIGHTCON]{Currently at Lightconnect Inc.}
\thanks[LMC]{Currently at Lockheed-Martin Corporation}
\thanks[LZH]{Currently at Laser Zentrum Hannover}
\thanks[MRC]{Currently at Mission Research Corporation}
\thanks[NASAGFC]{Currently at NASA Goddard Space Flight Center}
\thanks[NSF]{Currently at National Science Foundation}
\thanks[RAL]{Currently at Rutherford Appleton Laboratory}
\thanks[RAY]{Currently at Raytheon Corporation}
\thanks[REO]{Currently at Research Electro-Optics Inc.}
\thanks[UC]{Currently at University of Chicago}
\thanks[SHEF]{Currently at University of Sheffield}
\thanks[SIEM]{Currently at Siemens AG}
\thanks[SHANGHAI]{Currently at Shanghai Astronomical Observatory}
\thanks[SLAC]{Currently at Stanford Linear Accelerator Center}
\thanks[SPECTRA]{Currently at Spectra Physics Corporation}
\thanks[UCLA]{Currently at University of California, Los Angeles}
\thanks[ZEISS]{Currently at Carl Zeiss GmbH}
\thanks[DUB]{Permanent address:  University College Dublin}
\thanks[IAPARIS]{Permanent address:  GReCO, Institut d'Astrophysique de Paris (CNRS)}
\thanks[ICRR]{Permanent address:  University of Tokyo, Institute for Cosmic Ray Research}

\ \bigskip
\begin{abstract}
For  17 days in August and September 2002, the LIGO and  GEO
interferometer gravitational wave detectors were operated in
coincidence  to  produce  their first  data  for  scientific
analysis.  Although the detectors were still far from  their
design  sensitivity levels, the data can be  used  to  place
better  upper  limits  on  the flux of  gravitational  waves
incident  on  the  earth than previous direct  measurements. This
paper describes the instruments and the data  in  some detail, as
a companion to analysis papers based on the first data.
\end{abstract}

\end{frontmatter}
\setcounter{footnote}{0}


\section{Introduction} A number of laboratories around the world
[TAMA\cite{1}, VIRGO\cite{2}, GEO\cite{3},\cite{4},
LIGO\cite{5,6}] are developing detectors for gravitational waves
using laser interferometers to sense the very small strains
anticipated from astrophysical sources. In a joint effort,  two of
these laboratories,  LIGO and  GEO\,600, have performed  their
first scientific observations. This note is intended to provide
greater detail  in the  description  of the detectors  themselves
as  a companion to papers   describing the data   analysis   and
astrophysical conclusions from this Science Run (designated S1).

Both GEO\,600 and LIGO use the principle of the Michelson
interferometer, with its high sensitivity to differential changes
$\Delta L = L_1 - L_2$ of the lengths $L$ in the two perpendicular
arm lengths $L_1$ and $L_2$, to detect strains of the order of
$h=\Delta L/L = 10^{-20}$ over a wide frequency range. The
required sensitivity of the interferometric readout is achieved
through the use of high circulating laser power (to improve the
shot-noise limited fringe resolution) and through techniques to
store the light in the interferometer arms (to increase the phase
shift due to a passing gravitational wave). The frequency range of
interest for these instruments lies in the audio band
($\sim$50-5000 Hz), leading to gravitational wavelengths
$\lambda_{\rm gw} = c/f_{\rm gw}$ of hundreds of km. Because
practical ground-based detectors are short compared to the
wavelength, long interferometer arms are chosen to increase the
sensitivity of the instrument. A vacuum system protects the beams
from variations in the light path due to air density fluctuations.
The test masses, which also serve as mirrors for the Michelson
interferometer, are suspended as pendulums and respond as free
masses above their $\sim$1 Hz resonant frequency. External
mechanical disturbances are suppressed through seismic isolation
systems, and the in-band intrinsic thermal noise is controlled via
careful choice of materials and construction techniques.

\subsection{LIGO} The LIGO Observatory construction started in 1994
at the LIGO  site  in  Hanford,  Washington, USA. Construction at
the Livingston, Louisiana USA site began a year later in  June
1995. The buildings and  4 km concrete arm-support foundations
were completed  in 1998.  The vacuum systems were completed  in
1999, and detector installation was substantially completed in
2000. The first operation  of  a LIGO interferometer took  place
in October 2000. This marked the initiation of the commissioning,
consisting of periods  of intense testing  and tuning   of
subsystems, separated by periods where the interferometers  were
run  as complete systems. These engineering runs were  primarily
intended to assess the progress toward full detector  operation.
However, they were also used to collect data that could  be  used
to test data handling, archiving and analysis software. Progress
through the commissioning phase has been steady, both in terms of
improving sensitivity and in terms of reliable operation with  a
reasonable duty cycle.  By summer 2002, the improvements had  been
sufficient that a short duration Science Run could  be expected to
achieve limits on the observations of gravitational waves that
would  be comparable to or  better  than  previous experimental
limits. Consequently, a two-week observation period was scheduled,
and other laboratories operating  interferometer detectors were
invited to join in simultaneous operation, as documented here.
Further progress in sensitivity has subsequently been achieved
through additional commissioning.

\subsection{GEO\,600} The construction of GEO\,600 started  in  1995  as
a German/British  collaboration on a site near  Hannover, Germany.
Because the  site constrained  the length of the arms to 600m, an
advanced  optical layout  and novel techniques  for the suspension
systems were included in the detector  design. After the buildings
and the trenches were finished in  1997  the complete vacuum
system was installed and tested. The construction  phase  was
followed by the installation of the two suspended triangular mode
cleaners which have been operating reliably since 2000.  To gain
experience with the alignment and length control of long baseline
cavities the commissioning continued with the installation of a
1200 m long Fabry-Perot cavity formed by one interferometer arm
and the power recycling mirrors. To reduce the risk   of
contaminating or damaging   the expensive main interferometer
mirrors, lower grade test  mirrors suspended in steel wire slings
were used for the 1200 m cavity experiment and for   the
commissioning of  the power recycled Michelson interferometer
which started in summer 2001. A first engineering test  run  was
conducted in Jan 2002 in coincidence with  a  LIGO engineering
run.

The  installation  of  the  automatic alignment  system  for  the
Michelson interferometer and for the power-recycling cavity was a
key step towards a duty cycle of more than 98\% which was achieved
in  the  17  day S1 Science Run. The instrument ran as a
power-recycled folded-arm Michelson for S1; commissioning of
signal recycling started after S1 and  is expected  to bring the
GEO detector a significant step closer to its design sensitivity.

\section{Purpose of the S1 Science Run} The  primary goal of the
S1 run was to collect a significant body of  data  to  analyze for
gravitational  waves.   Although  the sensitivity  of all the
instruments was far from the design  goal and  the  relatively
short run time  made  it  unlikely  that  a positive  detection
would be made, it was  expected  that  upper limits could be
derived from the data that would be comparable to or better than
previous gravitational wave observations. Furthermore, the
analysis provided the opportunity  to  test  the methodologies
with  real  gravitational  wave  detector   data. Estimates  of
sensitivity for gravitational wave interferometers have  almost
always  been based on the  assumption  of  Gaussian noise.   While
this is a good point of departure for many of the limiting noise
sources (e.g., shot noise or thermal noise),  many  others (e.g.,
seismic noise) are not expected to be  so  well-behaved. Thus,
letting the data analysis confront the behavior  of  real noise as
early as possible is crucial to developing and  testing the
analysis techniques.

Other  goals for S1 were aimed at improving our understanding  of
the detectors and their operation.  These include:

\begin{enumerate}
\item Investigating the factors that influence duty cycle for the
     interferometers. Long periods of operation with stable conditions
     are   important  for  understanding  the  causes   for   the
     interferometers to lose `lock' (loss of resonance condition for
     light in the interferometer cavities and consequent loss  of
     linear operation of the sensing system)

\item Characterization of drifts in alignment and optimization of
     the alignment control systems

\item Testing and optimization of on-line monitoring tools  for
     assessing performance and maintaining high sensitivity

\item Training  and  practice  for  instrument  operators  and
     scientific monitors.
\end{enumerate}

This   paper  provides  a  description  of  the  LIGO   and   GEO
interferometers as they were used in the S1 run.\footnote{A
shorter period of simultaneous observations between TAMA, GEO, and
LIGO also took place during the period of this science run. That
effort will be documented elsewhere.} It is  intended as  a
companion to the data analysis papers based on  data  from this
run.   Because commissioning was still underway, many parts of the
detectors were not in their intended final  operational
configuration, and an important emphasis of this paper will be to
identify and highlight those differences.

\section{The LIGO detector array} The  LIGO detector array comprises
three interferometers  at  two sites.  The LIGO Livingston
Observatory (LLO) contains three main instrument bays at the
vertex and ends of the L-shaped  site  and houses  a  single
interferometer with 4 km long arms  (designated L1).   The  LIGO
Hanford Observatory (LHO)  has  two  additional experimental halls
at the midpoint in each arm which  enable  it to  accommodate two
interferometers, one with  4  km  long  arms (designated H1) and
one with 2 km arms (H2).  The orientation  of the  Hanford site
was chosen to be as closely  aligned (modulo $90^\circ$) to  the
Livingston site as possible, consistent  with  the   earth's
curvature and the need for the sites to be level; this maximizes
the common response to a signal, given the quadrupolar form of the
anticipated gravitational waves. The arms have an included angle
of $90.000^\circ$. The locations and orientations of the two LIGO
sites are given in Table~\ref{table:LIGO-where}.

\begin{table}[!hbp]
\caption{Location and orientation of the LIGO detectors. Note that
the Livingston Observatory is rotated by $\sim 90^\circ$ with
respect to the Hanford Observatory, when the earth's curvature is
taken into account.\medskip}\label{table:LIGO-where}

\begin{tabular}{l|c|c}
\hline
 LIGO Observatory:     &    Hanford                 &   Livingston \\
  \hline
Vertex Latitude     &       $46^\circ$ 27'18.5" N   &    $30^\circ$ 33' 46.4" N\\

Vertex   Longitude    &    $119^\circ$ 24' 27.6" W  &    $90^\circ$ 46' 27.3" W\\

Orientation of X arm  & $324.0^\circ$ (NW)          &   $252.3^\circ$ (WSW) \\

\end{tabular}
\end{table}

The  observatories have a support infrastructure of clean  rooms,
preparation   laboratories,  maintenance  shops,   and   computer
networking for control, data acquisition and analysis, as well as
offices for site staff and meeting spaces for larger gatherings.
The vacuum system can be divided into two main pieces:  the 4  km
beam  tube  arms (through which the laser beams pass between  the
vertex  and end test masses), and the vacuum chambers that  house
the suspended optics and associated equipment.  The vacuum tubing
for the arms\cite{13,14} is 1.2 m in diameter, fabricated from 3
mm thick 304L  stainless steel, processed to reduce the outgassing
to very low  levels (1 to $8 \times 10^{-14}\ \rm{mbar \cdot L
\cdot s^{-1} cm^{-2}}$). Expansion  bellows  are placed
periodically along the arms.  An extended bake at elevated
temperature was used to remove adsorbed water.  The tubing is
supported by a ground-level concrete  slab, protected   by  a
concrete  cover, and  aligned to centimeter accuracy\cite{15}. Ion
pumps at 2 km intervals and liquid nitrogen cooled cryogenic traps
where the arms enter the buildings at the vertex, end, and
midstations maintain  the base pressure in the arms between
$10^{-8}$ and $10^{-9}$ mbar,  with the residual  gas being mainly
hydrogen. This pressure is sufficient to  put the residual gas
scintillation well below the LIGO design sensitivity.

The   seismic   isolation   system,  test   masses,   and   other
interferometer  optics  are  housed in  vacuum  chambers  at  the
vertex, mid-stations (at Hanford), and end-stations.  Large  gate
valves  where the beam tubes enter the buildings allow the vacuum
chambers  to  be  isolated from the beam  tubes  and  brought  to
atmospheric  pressure  for  work on the  suspended  optics  while
maintaining  the  vacuum in the 4 km arms.  The  vacuum  chambers
have  large  doors  to  aid in the access to  install  and  align
optics.   When the chambers are at atmospheric pressure they  are
purged  continuously with clean (Class 10)  dry  air.  They  have
numerous viewports (for laser beams to enter and exit the  vacuum
system   and   for  video  camera  monitoring  of  the   interior
components)  and  electrical feedthroughs.   The  pumping  system
includes  roughing pumps with Roots blowers, and hydrocarbon-free
turbopumps.  Only  ion  pumps and cryogenic traps are used when
the interferometers  are operating.  The vacuum chambers are fully
instrumented  with  gauges and residual gas analyzers; pressures
range  between $4\times10^{-8}$ and $3\times 10^{-9}$ mbar. All
materials used in  the vacuum  chambers  and for the installed
detector equipment  are carefully  processed and screened to
minimize the  amount  of hydrocarbons  introduced into the vacuum
system as a  precaution against mirror contamination\cite{16}.

\begin{figure}
\caption{Schematic layout of a LIGO interferometer.}
\label{figure:LIGO-layout}
\begin{center}
\includegraphics[ width=0.8\textwidth]{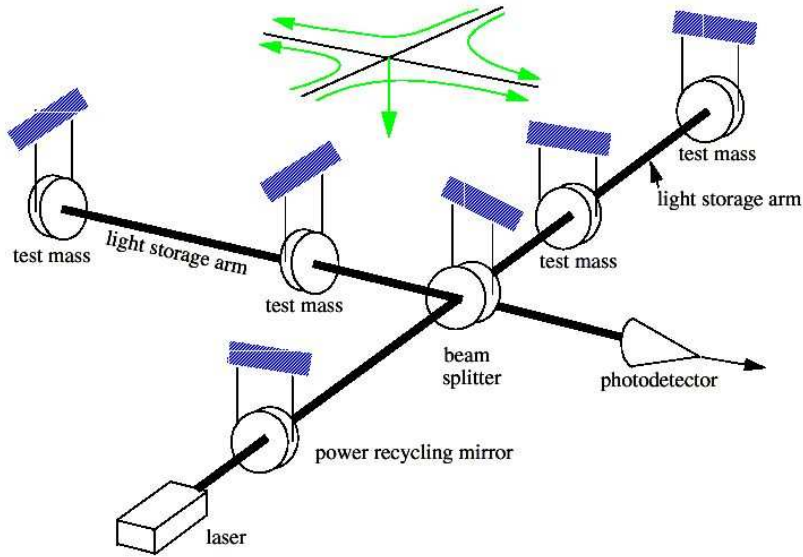}
\end{center}
\end{figure}

The basic optical configuration of each LIGO detector is that of a
power-recycled Michelson interferometer with resonant arm
cavities, shown in Figure \ref{figure:LIGO-layout}.  Gravitational
waves produce strains in space. The arm cavity mirrors serve as
the inertial test bodies (test masses), which move in response to
these strains. For example a sinusoidal wave incident on the plane
of the interferometer will cause an apparent shortening of the
optical path along one arm and a lengthening along the other arm,
and this process reverses half a cycle later in the signal
evolution. Laser light is incident from the bottom-left on the
beamsplitter, which divides it and sends it to low-loss cavities
in the arms.  The transmission of the input mirror in each cavity
is much larger than the losses in the cavity, and thus when the
cavities are on resonance, almost all of the light is returned to
the beamsplitter.  The beamsplitter is held so that the light
emerging from the antisymmetric port of the interferometer (right)
is at a minimum, and almost all of the light is reflected back
toward the laser. The power-recycling mirror forms a resonant
optical cavity with the interferometer, causing a build-up of
power in the recycling cavity.  The arm cavity mirrors serve as
the inertial test bodies (test masses), moving in response to the
gravitational wave.

\subsection{Laser}
\begin{figure}
\caption{Simplified schematic of laser stabilization. EOM:
Electro-Optic Modulator; AOM: Acousto-Optic Modulator; VCO:
Voltage Controlled Oscillator; PD: Photo Diode; PMC: Pre-Mode
Cleaner; IOO: Input Optics; LSC: Length Sensing/Control system}
\label{figure:LIGO-laser}
\begin{center}
\includegraphics[ width=1.0\textwidth]{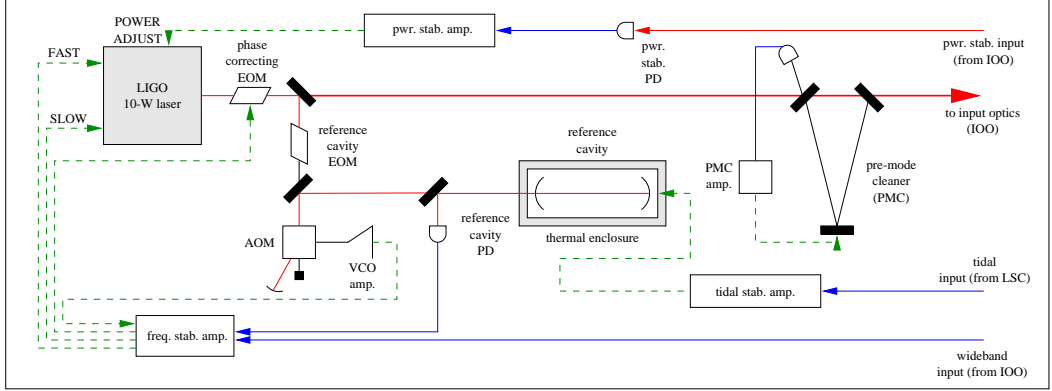}
\end{center}
\end{figure}

Each interferometer is illuminated with a continuous-wave
Nd:YAG laser operating in the $\rm TEM_{00}$ Gaussian spatial mode
at 1064 nm, and capable of 10 W output power\cite{17}.  A small
portion of the beam is used to stabilize the laser frequency using
a reference cavity in an auxiliary vacuum chamber (Figure
\ref{figure:LIGO-laser}).  The beam going to the reference cavity
is double-passed through an acousto-optic modulator driven by a
voltage controlled oscillator; this allows an offset between the
laser frequency and the reference cavity frequency to permit the
laser to follow the arm cavity length change due to tidal strains.
This initial level of stabilization is at $0.1\ \rm{Hz/Hz^{1/2}}$
or better in the gravitational wave band. The main portion of the
beam is passed through a ~45 cm path length triangular cavity to
strip off non-$\rm TEM_{00}$ light and to provide passive
filtering of the laser intensity noise with a pole frequency of
1.5 MHz (0.5 MHz at Livingston for the S1 run). An intensity noise
control system designed to reduce relative intensity fluctuations
below $10^{-8}\ {\rm Hz^{-1/2}}$ was only partially implemented
during the S1 run, leaving the intensity noise at approximately
$10^{-7}\ {\rm Hz^{-1/2}}$ level.  Electro-optic modulators
impress radio-frequency sidebands on the light at 24.5 and 33 MHz
(29.5 and 26.7 MHz for H2) for sensing respectively the
interferometer, and suspended mode cleaner\cite{18} degrees of
freedom. The design for the LIGO interferometers is for ~8 W to be
incident on the mode cleaner. However, the commissioning of the
instrument for high input power was not completed at the time of
S1, and the powers incident on the mode cleaner had been adjusted
(through the use of attenuators and reduced laser power) to
approximately 1 W for H1 and L1 and approximately 6 W for H2.

\subsection{Input  Optics}After the laser beam enters the vacuum  system,
it passes  through a set of input optics to condition it  before
it passes  to  the main interferometer.  First, it passes through
a mode cleaner -- a $\sim$24 m path length triangular ring cavity
with a finesse of $\sim$1350, formed from separately suspended
mirrors. This cavity stabilizes the beam size, position and
pointing.  It also blocks the 33 (or 26.7) MHz sidebands, but
transmits the 24.5 (or 29.5)  MHz  sidebands used for the
interferometer sensing, which are  at  multiples of the mode
cleaner free spectral range.  In addition, it serves  as  an
auxiliary reference for the laser frequency control servo,
reducing frequency noise  in the transmitted laser light to the
$10^{-3}\ \rm Hz/Hz^{1/2} $ level for  the  S1 run parameters.
After the mode cleaner, the beam passes through a Faraday
isolator, which diverts light returning from the interferometer
onto a photodetector.  This prevents the returning light from
reaching the laser and causing excess noise, and allows the
common-mode motions of the test-mass mirrors to be sensed.
Finally, the beam passes through an off-axis telescope formed  by
three suspended mirrors, which expands the beam to match the
$\sim$4 cm ($1/e^2$ radius in power) mode of the arm cavities.

\subsection{Interferometer  Optics}The main interferometer optics\cite{19,20}
are fabricated  from high-purity fused silica, 25 cm in diameter
and 10 cm thick (except the beamsplitter which is 4 cm thick).
Radii of  curvature  of the cavity optics are chosen so that  the
arm cavities have a stability $g = (1-L/R_1)(1-L/R_2)$ ($L$ is the
cavity length and $R_{\rm n}$ are the radii of curvature of the
two cavity mirrors) of 0.33 (H1 and  L1) or 0.67 (H2), to minimize
the excitation of higher order transverse modes
 by separating them in frequency from the laser frequency and its
 RF modulation sidebands. The surface figure accuracy of the  polished optic is better
than 1 nm; the coatings  have  a thickness uniformity that holds
their contribution to the apparent surface flatness negligible.
The coatings have a power absorption less than 1 ppm and scatter
less than 70 ppm. All optics  are wedged (typically about 2
degrees) to reduce the possibility of stray reflections
interfering with the main beam and   to give access to samples of
the  light inside the interferometer. Transmission of the input
mirrors  to the  arm cavities  is 2.7\% and the  end mirrors have
a transmission of approximately 12 ppm, to give an arm cavity pole
frequency  of ~85  Hz  (~170  Hz  for H2).   The beamsplitter
reflectivity was specified as $50\pm 0.5$\%.   The recycling
mirror transmission is also 2.7\%, to give a design recycling
factor (or increase in the circulating power) of $\sim$50 for the
optics as designed and at full power.

During  the S1 run, the low light input power led to the  optical
configurations of the three interferometers operating  away  from
their design point.  At full operating power, absorption of light
in  the  substrate and coating of the input mirrors for  the  arm
cavities is expected to create significant thermal lensing. As  a
result,  the  curvature of the recycling mirrors was figured to
compensate for this anticipated thermal lensing. Since  the
incident laser power in the H1 and L1 interferometers was
significantly under the design level, the lack  of  thermal
lensing  makes the recycling cavities slightly unstable for the
modulation sideband light. This has little effect  on  the carrier
recycling gain (since the carrier spatial mode is stabilized  by
resonance in the arm cavities) but reduces the transmission of
sideband light to the antisymmetric port by more than  a factor of
10. This further reduces the main differential arm length
sensitivity in the  high frequency region where shot noise is
expected to be dominant.  In the case of the 2 km interferometer
H2, although it was  receiving nearly the design input laser
power,  an out-of-specification anti-reflection coating on the
input mirror of one arm caused excess  loss  in the recycling
cavity and reduced the recycling gain  for the carrier by more
than a factor of two. As a result it also  did not  develop the
required thermal lens and its transmission of sidebands to the
dark port was also degraded by a similar factor. These limitations
contributed to the relatively high noise  level  of the
instruments seen in the sensing-noise limited regime (f $>$ 200
Hz).

\subsection{Suspensions}
\begin{figure}
\caption{LIGO Suspension} \label{figure:LIGO-suspension}\medskip
\begin{center}
\includegraphics[width=0.5\textwidth]{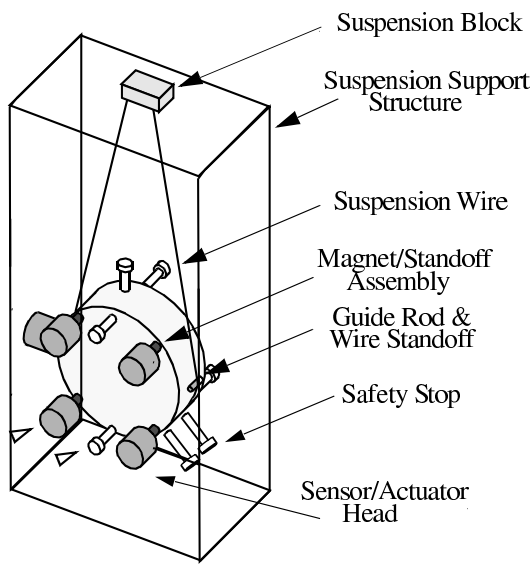}
\end{center}
\end{figure}

Each interferometer optic is suspended as a
pendulum from   vibration-isolated   platforms   to   attenuate
external disturbances   in  the  gravitational  wave  band; see
Figure \ref{figure:LIGO-suspension} for a schematic drawing. The
suspension fiber is a steel piano wire, loaded at approximately
40\%  of  its yield stress, passing under the optic as  a  simple
loop.   Small, notched glass rods are glued to the side  of  the
optic  a  few millimeters above the center of mass to define  the
suspension point  and minimize frictional  losses. The  normal
modes of the test mass optic suspension are approximately 0.74 Hz
(pendulum mode), 0.5 Hz (yaw mode), 0.6 Hz (pitch  mode),  12  Hz
(bounce mode), 18 Hz (roll mode) and multiples of 345 Hz (violin
mode). Thermal noise is managed in interferometric gravitational
wave detectors by placing resonances above or below the detection
band when possible, and by choosing materials and assembly
techniques which yield high resonance
$Q$'s\cite{saulson-fundamentals}. This gathers the thermal noise
power into a narrow band and lowers the values on either side of
the resonance. In the case of the suspensions, high resonance
$Q$'s (measured to be typically 2 to 4$\times 10^5$) in all
suspension modes yield a negligible level of off-resonance thermal
noise for the S1 sensitivity.

The  suspension  system  also provides  the  means  for  applying
control  forces  and torques to align the mirrors  and  hold  the
interferometer  in  resonance.  Four small  Nd:Fe:B  magnets  are
attached to the back of the mirror using aluminum stand-offs  and
a  vacuum compatible epoxy, with alternating polarities to reduce
coupling   to  environmental  magnetic  fields.   The  suspension
structure supports voice coils on ceramic forms near the  magnets
to  produce control forces. Each of these assemblies also
incorporates an LED/photodiode  pair  arranged  so  that  the
magnet   partially obstructs  the  path  between them  (a
``shadow  sensor").   This provides  a  read-out of the
longitudinal position of the  magnet with  a  noise  level  of
approximately  $10^{-10}\ \rm m/Hz^{1/2}$ in the gravitational
wave band.  Similar magnets are attached to the sides  of the
optic and a shadow sensor/voice coil assembly acts on one of these
to damp sideways motion.

The  magnet  and coil actuators are driven by several sensors, via
servo controllers, allowing control of their positions and
orientations with respect to both the local structures and the
globally-measured lengths and angles. Local damping  of  the modes
of  the suspension  is provided by feeding appropriately filtered
and mixed signals from the  shadow sensors to the coils to create
a damping force near the  pendulum frequencies. Signals  from
interferometric-based wavefront sensors  and optical levers
(described below) are also applied to maintain the pointing   of
the  test mass. Lastly, the interferometer length signals are
applied to acquire resonance and hold the operating lengths for
the interferometer to  within $\sim 10^{-13}$ m rms. The
suspension controllers, which combine and filter these signals
appropriately, were of two styles during S1: an original analog
system with some digital gain and filter controls (for  H2 and
L1),  and  a system with the signal processing performed
digitally\cite{21} (H1). In all cases,  a significant
low-frequency noise contributor was the final amplifier, which for
S1 had  to  deliver stronger control forces than those expected
for the  final configuration. This then compromised the
gravitational wave-band performance.

Thermal noise internal to the mirrors is minimized by maintaining
high  $Q$'s  in  all the internal modes. The fused silica internal
losses  are  anticipated  to make the  dominant contribution  to
thermal  noise.  However, the dielectric coating on  the  mirror
will  also contribute noticeably because of its proximity to  the
beam\cite{22}. The attachments to the mirror for the suspension
and the magnets  can  degrade the individual modal $Q$'s  but,
because of their distance from the front surface of the optic,
their effect on thermal noise is negligible. In-situ measurements
of Q's typically range from $2\times 10^5$  to  $1.6 \times 10^7$,
depending on the mode. Calculations indicate that the  thermal
noise  is  near the design  goal and  thus negligible for the
S1-run sensitivity.

\subsection{Seismic Isolation}The vibration isolation systems are four
layer passive isolation stacks\cite{23}. The final stage in each
vacuum chamber is  an  aluminum  optical table that holds the
optic suspensions. Each  optical table is supported by four legs.
Each leg consists of  a  series of three heavy stainless steel
cylinders, supported by  coil springs made with phosphor bronze
tubing containing inner constrained  layers which  are  sealed
from  the vacuum  via electron-beam  welding.   The transfer
function of ground motion to table motion shows a series of  broad
peaks between 1.5 and 12 Hz, representing  the  normal modes
(typical  $Q\sim 10-30$) of the masses moving  on  the  springs,
followed  by  a  steep falloff above the highest resonance.   The
total  attenuation reaches a value of $10^6$ by about  50  Hz. The
high  $Q$'s  of the resonances in the 1.5 to 12 Hz band presents a
particular problem at  LLO, where they amplify anthropogenic
ground  noise in  this frequency range, and cause difficulties in
locking during daylight hours. A planned six degree-of-freedom
external active isolation system to cope with this excitation was
not in place during S1. The support points for the seismic
isolation stack penetrate the vacuum chamber through bellows that
decouple the seismic isolation stack from vacuum chamber
vibrations  and drift. External coarse actuators at the support
points permit translations and rotations to minimize the control
forces that are needed  to align  the  optics  during  the initial
installation and to compensate for any long-term settling.

In  addition,  the systems at the ends of the arms  are  equipped
with a fine actuator aligned with the arm that can translate  the
entire assembly (seismic isolation stack and optic suspension) by
approximately   $\pm 90 \rm\mu m$ over the frequency range from DC
to $\sim$~10~Hz. This system is used during the interferometer
operation to compensate for earth tides, using a simple predictive
model and a very slow feedback from the differential and  common
mode arm length controls.  At LLO, an additional microseismic
feed-forward system\cite{24} was used to reduce the length
fluctuations of the arms at  the microseismic frequency
(approximately 0.16~Hz). Also, the L1 detector's fine actuators
were used together with seismometers in a beam-direction active
seismic isolation system at each test mass chamber, which reduced
seismic excitation of the most troublesome stack modes by a factor
of $\sim$5.

\subsection{Length  and Angle Control}There are four longitudinal degrees
of freedom  that must be held to allow the interferometer to
function: the two arm lengths are held at the Fabry-Perot cavity
resonance condition, the beamsplitter position is set to maintain
the light intensity minimum at the antisymmetric port and the
recycling mirror position is positioned to meet the resonance
condition in the recycling cavity. These lengths are sensed using
RF phase modulation sidebands on the incident light in an
extension\cite{25} of the Pound-Drever-Hall technique. The
modulation frequency was chosen so that  the phase modulation
sidebands are nearly antiresonant in the  arm cavities; the
carrier light is strongly overcoupled so that ~0.97 of  the light
is reflected  on resonance,  and  it receives a $\pi$ phase shift
on reflection.  By making the recycling round trip  cavity length
an odd number of RF half-wavelengths, the recycling cavity can be
simultaneously resonant for the carrier and sidebands. A small
length asymmetry ($\sim$30~cm) is introduced between the
beamsplitter and the two input test masses to couple the sideband
light out the dark port.

\begin{figure}
\caption{Schematic drawing of a LIGO interferometer showing laser,
input light mode cleaner, and the locations of the photodiodes
($\rm S_{xxx}$) used to sense and control the resonance
conditions. $L_1$ and $L_2$ are the arm cavity lengths; a
gravitational wave produces a differential signal of the form
$(L_1-L_2)$, and $(L_1+L_2)$ is a sensitive measure of the laser
frequency noise. The Michelson degrees of freedom are differential
$(l_1-l_2)$ and common-mode $(l_1+l_2)$, the latter measured with
respect to the recycling mirror. PBS: Polarizing  Beam Splitter.
AOM: Acousto-Optic Modulator. PC: Pockels Cell. VCO: Voltage
Controlled Oscillator.  $\rm S_{mc}$: Signal, Mode Cleaner. $\rm
S_{ref}$: Signal, Reference Cavity. $\rm S_{refl}$: Signal,
reflected light. FI: Faraday Isolator. PRM: Power Recycling
Mirror. $\rm S_{prc}$: Signal, Power Recycling Cavity. $\rm
S_{anti}$: Signal, Antisymmetric Port. BS: Beamsplitter. ITM:
Input Test Mass. ETM: End Test Mass.}
\label{figure:LIGO-readout}\medskip
\begin{center}
\includegraphics[width=0.85\textwidth]{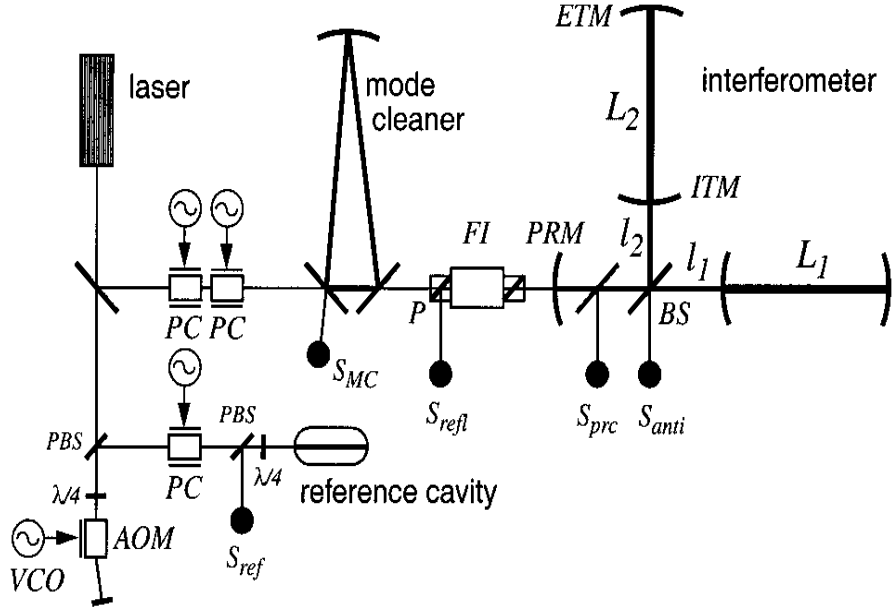}
\end{center}
\end{figure}

Three  interferometer  output  beams  (Figure
\ref{figure:LIGO-readout})  are  used   to determine the
longitudinal degrees of freedom\cite{26}, which are best thought
of as two differential motions (arm cavities  or strain readout,
and the Michelson), and two common-mode motions (common mode
`breathing' of the arm cavities, and of the power recycling
cavity).  A photodiode  signal  at  the  antisymmetric port  is
demodulated with the $90^\circ$ quadrature of the modulation drive
to give  a signal  proportional to the difference  in arm cavity
lengths (differential arm length).  A second photodiode monitors
the light reflected from the recycling mirror (separated from the
incident beam by a Faraday isolator); it is demodulated in  phase
with  the modulation  drive and is primarily  sensitive to  the
average  of length  of  the two arm cavities  (common mode  arm
length). The third photodiode monitors  light  from inside  the
recycling cavity, picked  off from the back
(anti-reflection-coated) side of  the beamsplitter with  the aid
of the  small  wedge  angle in the substrate.  The in-phase signal
is primarily  sensitive to the recycling  cavity length, while the
quadrature phase is sensitive to  the  Michelson path difference
from the beamsplitter to  the input   test masses. One major
deviation  from   the final interferometer design during S1 was
that attenuators were placed in  front  of the antisymmetric port
photodiodes  on  all three interferometers, reducing  the
effective  power  used  in each interferometer to about 50 mW
instead of the 6 W nominal value. These attenuators protected  the
photodiodes  from  saturation, and possible damage, during  the
commissioning  phase  before the complete mirror angular controls
were  implemented  and when large fluctuations in the power on the
photodiodes were present. This had  a particularly significant
impact on the performance in the high  frequency region (above a
few hundred Hz),  where  the low effective  light level combined
with the reduced sideband efficiency noted above to cause the
noise to be well above the design level.

The  signals from these three photodiodes, appropriately
demodulated and filtered,  are used to control the lengths   and
hold the interferometer in resonance. The high frequency portion
of the reflected photodiode signal $\rm S_{refl}$ is fed back (via
an analog path at Hanford for S1, digital  at Livingston) to the
mode cleaner and laser to stabilize the  input laser frequency to
the average arm length. The signals $\rm S_{prc}$ and $\rm
S_{anti}$ from  the other two photodiodes are used to control the
positions of  the interferometer  optics. The demodulated signals
from  all three photodiodes are whitened with an analog filter,
digitized with  a 16  bit  ADC operating at 16384 samples per
second, and digitally filtered  with  the inverse of the analog
whitening filter  to return  them to their full dynamic range.  A
dedicated real-time signal processor combines these error signals
via a matrix (whose coefficients  are adjusted  in real  time
during the  lock acquisition process)  to form appropriate control
signals, filters them,  and sends  the results to combinations of
optics to control the interferometer. It also passes the
photodiode signals (error signals) and the feedback signals to the
data acquisition system. The  flexibility of  the digital control
system  to respond  in changes  to  the interferometer  response
function during  the `locking'  process as a function of sensed
light levels,  and  to allow specialized filters to be implemented
on the fly, has  been crucial  to the ability to acquire lock on
the interferometers\cite{27}, to aid in the commissioning,  and
ultimately to suppress noise in the control systems.

As  noted  above,  an  ensemble of optical levers  and  wavefront
sensors  is designed to sense and control the angular degrees  of
freedom  of  the  suspended  optics  in  the  main  part  of  the
interferometer\cite{28,29}.  Each large (25 cm) optic is equipped
with an optical  lever, consisting of a fiber-coupled diode laser
and a (position  sensitive) quadrant photodiode, which is intended
to hold  the  optic  stable  while the interferometer is unlocked.
These  components are mounted on piers outside the vacuum  system
and  operate  through viewports at distances between  1  and  25
meters from the optic; their long-term stability and independence
from the interferometric sensing system allows a manual alignment
to be maintained continuously.

The  full instrument design includes a wavefront sensing  control
system  to  optimize the alignment during operation. Quadrant
photodiodes  are  placed  at  the  output  ports  of the
interferometer, in the near field and (via telescopes) in the far
field.  The  photocurrents  are demodulated  as  for  the length
control system, and sums and differences can be formed to develop
a complete set of alignment information which is then used to
control the mirror angles, using the suspension actuators.
However, at the time of the S1 run, this  system was only
partially commissioned, and only the  mode cleaner  and  two
degrees  of freedom  of  the interferometer, the differential
pitch and yaw of the end test masses (cavity end mirrors), were
controlled by wavefront sensors.  As  an interim measure,  the
incomplete wavefront sensing was complemented  by signals  from
the optical levers during operation. However,  the optical lever
angular sensing noise is much greater than that for the  wavefront
sensors. Even after careful control-law  shaping, the optical
levers remained one of the principal contributors  to the
low-frequency noise of the instrument for S1.

Baffles  to capture stray light are placed along the  4  km  beam
tubes,  and at specific places near the optics inside the  vacuum
chambers,  to reduce the possibility of a scattered beam  or  one
from an intentional wedge in the optics from recombining with the
main  beams. Some of the baffles for the final installation  were
not  in place for the S1 run, but calculations indicate that this
should  not have been a source of noise at our present  level  of
sensitivity.

\begin{figure}
\caption{A frequency-domain model of the noise sources at the time
of the S1 run for the Livingston (L1) detector. The noise sources,
discussed in the text, are assumed to add in quadrature. The
actual noise curve is also shown, along with the performance
expected for the instrument when working at the design level}
\label{figure:LIGO-noise-model}
\begin{center}
\medskip
\includegraphics[width=.95\textwidth]{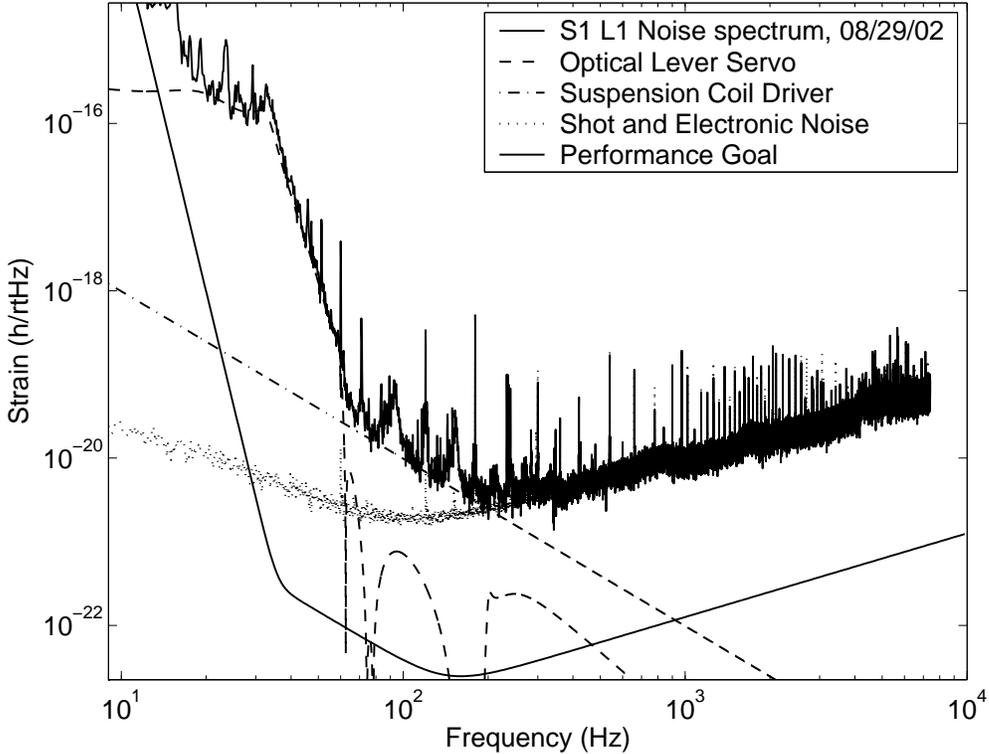}
\end{center}
\end{figure}

A noise model of the instruments summarizes the limits to the
performance of the interferometer at the time of this science run.
The model for the Livingston detector is shown in Figure
\ref{figure:LIGO-noise-model}. In general, the contributions are
evaluated by measuring a source term (e.g., laser frequency
noise), and measuring a coupling function (e.g., the transfer
function from an intentional frequency modulation to the response
in the strain channel), and then multiplying these two together to
make a prediction. In some cases, analytical models are used (for
example the mechanical $Q$ of the suspension systems is measured
and then used in a model of the thermal noise contribution). For
this model, all the terms are considered to be independent, and
the noises are added as the square root of the sum of the squares.
Many sources of noise have been modelled; this figure only shows
those that limit the present performance. The model explains the
overall instrument noise performance well, and subsequent
commissioning efforts have shown that reductions in the leading
noise terms also leads to the anticipated reduction in the overall
instrument noise.

\subsection{Simulations}In addition to subsystem dynamics and control
models, two  simulations  played a significant role  in  the
design  and commissioning of the LIGO detectors.  The first is  an
FFT-based optical   propagation  code\cite{30}  that  models  the
power-recycled Michelson  interferometer with Fabry Perot arms.
This  code  was used to develop the specifications for the
interferometer optics, and  has  been  used  in comparisons with
commissioning  data  to evaluate the performance of the optics as
installed. The second simulation is an end-to-end time-domain
simulation  of the  LIGO interferometers\cite{31}.  This model
includes a modal-based optical propagation,  accurate  modeling of
the electronic feedback, simplified  models  for the suspension
systems,  and typical noise  inputs. This model proved to  be
invaluable  in developing the lock acquisition software.

\subsection{Environmental  Monitoring}A  system  of  auxiliary   sensors is
installed  at  each  LIGO site to monitor possible environmental
disturbances.  The  Physics  Environment  Monitor system   (PEM)
contains  seismometers  and tiltmeters to monitor low  frequency
ground  disturbances, accelerometers and microphones  to  monitor
higher   frequency  mechanical disturbances,  magnetometers   to
monitor  magnetic fields that might affect the test  masses,  and
monitors of the line power. Sensors are present in all buildings
and  near  all  key sensitive components. They have been used to
e.g., help identify sources of acoustic and electromagnetic
coupling, and to help design improvements to the apparatus; as the
instrument sensitivity improves, they will be used as veto signals
in the astrophysical analyses. Planned cosmic ray detectors and rf
monitors were not operational at the time of the S1  run.

\subsection{Control   and   Data   Systems}Supervisory   control    of the
interferometers is accomplished using EPICS (Experimental Physics
and   Industrial   Control  System\cite{32}).    EPICS establishes
a communications protocol within a non-hierarchical computer
network and  provides  an operator interface from networked
workstations located  in  the  control room.  Processors
distributed  in  all electronics racks can modify amplifier gains,
offsets, filtering, on/off  controls,  etc.,  allowing  either
manual  or  automated (scripted)  control of the state of the
electronics.   The  EPICS processors  also  interface  to
analog-to-digital  converters  to provide  monitors for the
electronics inputs and  outputs.   Each interferometer  has
approximately 5000 EPICS  variables  (either control  or  monitor
points).  EPICS  also  provides  tools  for capturing  and
restoring the state of the instrument  to  ensure that  this
complex instrument can be reliably brought to a  known
configuration.

The data acquisition and control system collects signals from the
interferometer and from the environment, and delivers signals for
the   length   and  angle  controls.  VME-based  converters   and
processors  are used, and acquisition systems are placed in the
vertex building and the mid- and end-stations. Analog signals are
digitized with 16-bit resolution. Fiber optics are used to link
the instrument  racks together, and a shared memory approach
allows data  to be collected and shared by a number of systems
over  the multi-km   distances.   The  data are collected with
16-bit resolution.  The  complete data are formatted into the
standard data  `Frames'  (a  format  used by all  of  the
interferometric gravitational  wave  community)  and initially are
stored   on spinning  media  for a quick `look-back' buffer of
roughly  two weeks.   All  data  are archived to tapes for later
analysis. Reduced  data sets also in the standard Frame format,
configured for  a  given  science run, are produced as well; these
serve most analysis needs.

It is important that the data acquisition system accurately
time-stamp the data it records.   The  fundamental timing for both
sites is derived  from GPS receivers located at each building
(vertex, mid and end).   A $2^{22}$  Hz (approximately 4.2 MHz)
clock signal is generated from the GPS  as  well  as  a  1
pulse-per-second synchronization signal. Together these are used
to synchronize the data collected by  the various processors. Ramp
signals are used to monitor any timing errors and alarms are set
for the operators. This monitoring has proven useful during S1. It
showed that the timing was subject to jumps (typically 10's of
milliseconds, but sometimes larger) when the length control system
processors were rebooted, with the consequence that some S1 data
had to be eliminated from some analyses because of uncertain
timing (These timing jumps have since been cured, and a redundant
and independent atomic clock reference is being implemented for
the future).

\subsection{Diagnostics  and  Monitoring}Two  closely  related  systems,
the Global  Diagnostic  System and the  Data  Monitoring Tool,
provide the instrument operators and scientific monitors with
tools for evaluating interferometer performance both during
commissioning  and  scientific running\cite{33}. The Global
Diagnostic System (or GDS) can  access data from any signal
collected by the  data acquisition system, including test-point
signals that can be stored for post-analysis if indicated. It can
display the time series and the power spectrum for individual
signals, and the transfer function and coherence for pairs of
signals.  The GDS also has the ability to apply arbitrary-waveform
excitations to various test points within the interferometer
control systems. These can be used  to  measure transfer functions
through stimulus-response testing. The data can be  filtered,
decimated, calibrated, stored and  recalled  for comparisons.

The   Data  Monitoring  Tool  (or DMT)  is  a  package  of
software components running on a set of processors on a dedicated
network. A  high-speed connection to the data acquisition system
makes the full  data  set  available with only a one-second
latency.   The emphasis  in  the  DMT  is  on  relatively  simple
measures   of instrument   performance  applied  to  the  full
data stream   in realtime.  Thus it can give the operators and
scientific monitors rapid  feedback about interferometer
performance.  These  include such measures as the non-stationarity
and burst-like behavior  of various  types  in  the interferometer
outputs, band-limited  rms amplitudes of interferometer outputs
and environmental  monitors, monitors   of  calibration  lines,
histograms  to  monitor   the gaussianity  of  the  data, and
real-time estimates  of  detector sensitivity  to  neutron star
binary inspiral  events.  The  DMT is also an important element of
the data analysis process,  analyzing the auxiliary channels for
veto  signals  in parallel with the strain channel analysis.

\subsection{Data   Analysis  System} To  analyze  the  large  volume  of
data generated,  LIGO  has developed the LIGO Data  Analysis
System (LDAS). The LDAS provides a distributed software
environment with scalable  hardware  configurations to provide the
computational needs  for  both on and off site data analysis. The
architectural design  of  the  system  is  based on  the concept
of  multiple concurrent data analysis pipelines in which data is
fed into  the pipeline as it is collected and proceeds down the
pipeline  where necessary signal analysis procedures are applied
depending on the particular  type  of analysis that is  being
carried out\cite{34}.

LDAS is complemented by the LIGO/LSC Algorithm Library, which is a
set of $\tt C$-language routines that can run under LDAS or be
used independently. They are carefully vetted to ensure that the
algorithms and results are correct.

The  LDAS distributed software environment is composed of roughly
12 modules called LDAS Application Programming Interfaces (APIs),
each  of  which  is a separate process under the  Unix operating
system. Each module is designed to carry out the multitude of
steps associated with each  unique  pipeline.  For example, one
module  has  computational elements  for reading and writing LIGO
channel data in the  Frame format, another module has
computational elements for  signal processing in the time or
frequency domain, and another  module has computational elements
necessary to perform parallel analysis across a cluster of tightly
networked CPUs\cite{35}. Upon completion of the  data analysis
pipeline, data products and results are stored to  disk  or
inserted  into  the LIGO relational database. The database  has
tables designed to capture results associated with detector
characterization, on line and off line astrophysical searches and
multi-detector analyses. The software can be  scaled to  run on a
wide variety of computing hardware.

During  the first LIGO Science Run, the LDAS at the  LIGO
Observatories  and data analysis centers located at  Caltech  and
MIT, LDAS at other institutions, and other
configuration-controlled computational systems were operated with
commissioning configurations of the hardware and software. The
software was in the late stage of beta development, having  a
complete set of modules.  The hardware systems consisted of  a
complement of servers  with tens  of terabytes of disk storage for
the raw data and the LIGO database, along with scaled down
computation centers with approximately 200 megaflops of aggregate
computational performance between them. In its  final
configuration, the LDAS hardware will include upgrades to the
current servers, and expanded high performance computation
clusters  with  over  two teraflops of aggregate  computational
performance. In addition, new tape storage systems will be put on
line which will provide adequate storage at the observatories for
six  months  of  local data and storage for all of  the data  at
Caltech; this is where the data for the multiple detectors are
brought together. The software is expected to double in
performance as we upgrade from beta versions to the first
completed version later this year. In addition, the software is
being adapted to support Grid Computing technology and security
protocols allowing for LIGO data analysis once the computational
Grid is deployed in the near future\cite{36}.

\subsection{LIGO  Data}The  full  data stream from each of the LIGO
interferometers consists of several thousand channels, recorded at
rates  from  1 Hz   to   16384  Hz  with  a  total  data  rate  of
5MB/s   per interferometer.   These channels include EPICS process
variables that  define  the  state  of  the  interferometer,
signals  from environmental  monitors,  signals from auxiliary
servos  in  the interferometer (for example, optical lever
signals), as  well  as the   main  gravitational  wave  signal.
For  the  servos   not operational  during  S1,  the corresponding
data  channels  were recorded,  but  of  course they contain  only
zeros.   The  non- gravitational wave data channels can be used in
a number of ways:

\begin{enumerate}
\item They can be used to determine the operational ``health" of
     the  interferometer (how well it was aligned, whether  large
     offsets were present in any servos, etc.).

\item They can be used to regress noise from the main
gravitational wave channel (for example, measurements of the laser
frequency noise can be used to correct the gravitational wave
channel to remove any residual effects from laser frequency noise
coupling to mismatches in the arms).

\item They can be used to veto non-gaussian noise in the
interferometer (for example seismometer data could be used to keep
noise from impulsive seismic disturbances from being
misinterpreted as a gravitational wave).

\end{enumerate}

At  this point in the commissioning, few of these techniques have
been explored and developed.  In part,
this  is  because  the  majority of  the  noise  sources  in  the
interferometer  at present are attributable to  electronic  noise
entering through imperfect tuning, and consequently, few  of  the
auxiliary   channels  are  expected  to  be   useful.   The   DMT
capabilities   to  perform  this  analysis  were   exercised   in
preparation and performed very well.

The main signal for the analysis to search for gravitational wave
signals  is  the  output of the photodiode at  the  antisymmetric
port,  demodulated in the quadrature phase at 24.5  MHz  (H1  and
L1),  or  29.5  MHz  (H2).  This analog  signal  is  amplified and
digitized.    An   analog  filter  whitens  the  signals before
digitization,  and a precise inverse of this filter `de-whitens'
the  signal in the digital domain, to best take advantage of  the
dynamic range and noise in the Analog-to Digital Converter (ADC).
Since  it  is  the error point in the servo control system  which
holds  the  differential arm length, its interpretation  requires
correction for the loop gain of the servo. This signal represents
the  phase  difference of the light from the two  arms,  filtered
only  by a roll-off at high frequencies because of the arm cavity
storage  time,  and while the interferometer is operating,  is  a
continuous measure of the differential strain between the  two
arms and thus potentially of gravitational wave signals.

\section{The GEO Detector} The  GEO Detector is situated at the
perimeter of an agricultural research  station  to  the south-east
of Hannover, Germany; see Table~\ref{table:GEO-where}.  The
buildings are intended to be just sufficient to accommodate  the
instrument and its acquisition and  control hardware.   Data
recording,  and much of the operation and on-line monitoring  of
the instrument, will be performed at the Max Planck Institute  in
downtown Hannover, once continuous science operation is underway.
A microwave link maintains a high-bandwidth dedicated connection
between the two.

\begin{table}[!hbp]\medskip
\caption{Location and orientation of the GEO\,600 detector. Note
that  the  arms  form an angle of  $94^\circ$  19'  53".   This
deviation from  perpendicular  has  negligible  effect  on   the
sensitivity.} \label{table:GEO-where}
\begin{tabular}{l|c}
\hline

Vertex Latitude  &  $52^\circ$ 14'
                   42.5" N \\
Vertex Longitude &  $9^\circ$ 48'
                   25.9" E\\
 Orientation of  North arm  & $334.1^\circ$
         (NNW)\\
 Orientation of   East arm  &  $68.4^\circ$
         (ENE)\\
\end{tabular}
\end{table}

One  central  building  (13 m $\times$ 8 m in size) and  two  end
buildings (6 m $\times$ 3 m)  accommodate  the  vacuum  chambers
(2 m tall, 1m  in diameter) in which the optical components are
suspended. In the central  building, nine vacuum chambers form a
cluster which can be  subdivided  into three sections to allow
mirror installation without  venting  the whole cluster. This
arrangement  allows a minimum of down-time for a  change  of  the
signal-recycling mirror  (to change  the   detector bandwidth). To
avoid fluctuations of the optical path caused by a time-varying
index  of  refraction,  all  light  paths  in   the interferometer
are in a high-vacuum system. For this purpose  GEO\,600  uses two
600 m long vacuum tubes of 60 cm diameter which are suspended  in
a  trench  under ground. A  novel  convoluted-tube design,
allowing a wall thickness of only 0.8 mm, was used to reduce
weight and cost of the stainless-steel vacuum tube.

The  whole vacuum system, except for the mode cleaner and  signal
recycling section, is pumped by four magnetically levitated turbo
pumps  with a pumping speed of 1000 l/s, each backed by a  scroll
pump  (25  $\rm m^3/h$). Due to the use of stainless steel  with a
low outgassing rate, a 2 day air bake at $200^\circ$C and a 5 day
vacuum bake at  $250^\circ$C,  a pressure of $1\times 10^{-8}$
mbar can be achieved in the tubes.  Large  gate  valves  allow the
beam tubes  to be temporarily  closed off and maintained under
vacuum whenever the instrument  vacuum  chambers are opened  for
installation work. Additional  dedicated  pumping systems  are
used  for  the mode cleaner  section  and  for  the
signal-recycling  section. The pressure in the vacuum chambers is
in the mid $10^{-8}$ mbar range. Great  care was taken to minimize
contamination of the all-metal vacuum  system  by  hydrocarbons.
For  this  reason  the seismic isolation  stacks,  which contain
silicone  elastomer  and other materials  containing hydrocarbons,
are sealed  by  bellows and pumped separately. Furthermore, the
light emitting diodes (LEDs), the  photodiodes and the feedback
coils used as `shadow' sensors and  actuators  in the pendulum
collocated damping and actuation systems are encapsulated in
glass.

The  buildings  of  GEO\,600 are split into  three  regions  with
different  cleanroom classes: the so-called gallery where  people
can  visit  and  staff can work with normal  clothes,  the  inner
section  which  has  a  cleanroom class of  1000  and  a  movable
cleanroom  tent  installed over open chambers  with  a  cleanroom
class 100.

\subsection{Suspension and Seismic Isolation}Two  different  types  of  seismic
isolation and suspension systems are implemented in GEO\,600. The
first one, used to isolate the mode cleaner optics, consists of a
double pendulum suspended  from a pre-isolated top-plate. To avoid
an excitation of the pendulum mode, four collocated control
systems measure the motion of the intermediate mass with respect
to a coil-holder arm which  is rigidly attached to the top plate,
and feed back  to  the mirror via a coil-and-magnet system.

The  seismic  isolation system used to isolate the  test  masses,
beamsplitter,   and  the  other  mirrors  of   the   Michelson
interferometer  consists of a triple pendulum\cite{9} suspended
from a pre-isolated platform.  Each pendulum chain consists of the
optic suspended  from  an intermediate mass which is in turn
suspended from an upper mass.  Two cantilever spring stages are
included in the pendulum design (in the support of the upper and
intermediate masses)  to reduce the coupling of seismic motion in
the vertical direction  to  the  mirror. As in the case of  the
mode  cleaner pendulums, collocated feedback systems are used to
damp  all  six degrees of freedom of the upper pendulum mass and
through cross-coupling  the  other  solid-body modes of the
multiple pendulum system. The control forces for the length and
alignment control are  applied from a reaction pendulum which
consists of a similar triple  pendulum suspended 3\,mm behind the
corresponding mirror. The  intermediate  mass  of the reaction
pendulum  carries coils which act on magnets glued to the
intermediate mass of the mirror triple  pendulum.  To  keep the
internal quality  factor of the mirrors  as high as possible, no
magnets are glued to the mirror itself,  but  electrostatic
feedback between the mirror and the lowest  mass  of the reaction
pendulum is used to apply feedback forces in the high frequency
control band.

\begin{figure} \caption{An outline
sketch of the test mass suspensions in GEO\,600. The test mass and
intermediate mass are made of fused-silica and are connected by 4
fused-silica fibres.  The reaction mass is also of fused-silica
and has the electrode pattern required to allow electrostatic
actuation forces to be applied to the test mass. The other masses
are fabricated from metal.  Two stages of vertical isolation are
provided in the form of cantilever mounted blade springs. Active
local damping is provided from a structure (not shown) held at the
upper mass level by the damping arms. The pendulum chains are
suspended from a structure which allows crude angular alignment of
the mirror, and this is in turn supported by 3 vibration isolating
legs.  Each leg consists of an active layer and a passive layer
which, to avoid  contamination of the vacuum are enclosed in steel
bellows. A flex-pivot is then required to provide rotational
compliance. The reaction chain is omitted from the test mass
suspensions in the end stations.} \label{figure:GEO-suspension}
\begin{center}
\medskip
\includegraphics[width=0.5\textwidth]{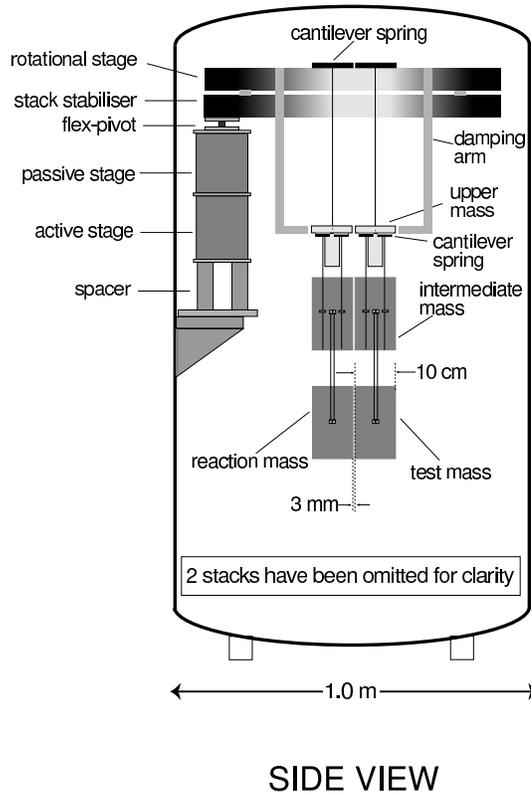}
\end{center}
\end{figure}

\subsection{Suspension}To  further minimize the mechanical  losses  and
thus internal  thermal  noise of the mirrors and the  pendulums,
the  lowest  pendulum stage consists  entirely  of fused silica;
see Figure \ref{figure:GEO-suspension}.   The mechanical  quality
factor $Q$ of fused-silica suspensions comparable in size has been
demonstrated\cite{37} to be greater than $2\times 10^7$.  Small
fused-silica pieces are attached  to the intermediate mass and to
the mirror itself by hydroxide-catalysis bonding\cite{38}. This
technique provides high-strength bonds and allows the  high
quality  factor to be maintained  and  therefore the thermal noise
to be kept low. Four fused-silica fibers of 270 $\mu$m diameter
each are  welded to these fused-silica pieces and support  the
mirrors.

\begin{figure}
\caption{Optical Layout of GEO\,600. A 12 W injection locked laser
system is filtered by two sequential mode cleaners and injected
into the dual (power and signal) recycled interferometer.  Only
power recycling was used for the S1 run. A folded light path is
used to increase the round-trip length of the interferometer arms
to 2400 m. An output mode cleaner will be used to spatially clean
the laser mode before it reaches the photodetector}
\label{figure:GEO-layout}
\begin{center}
\includegraphics[width=1.0\textwidth]{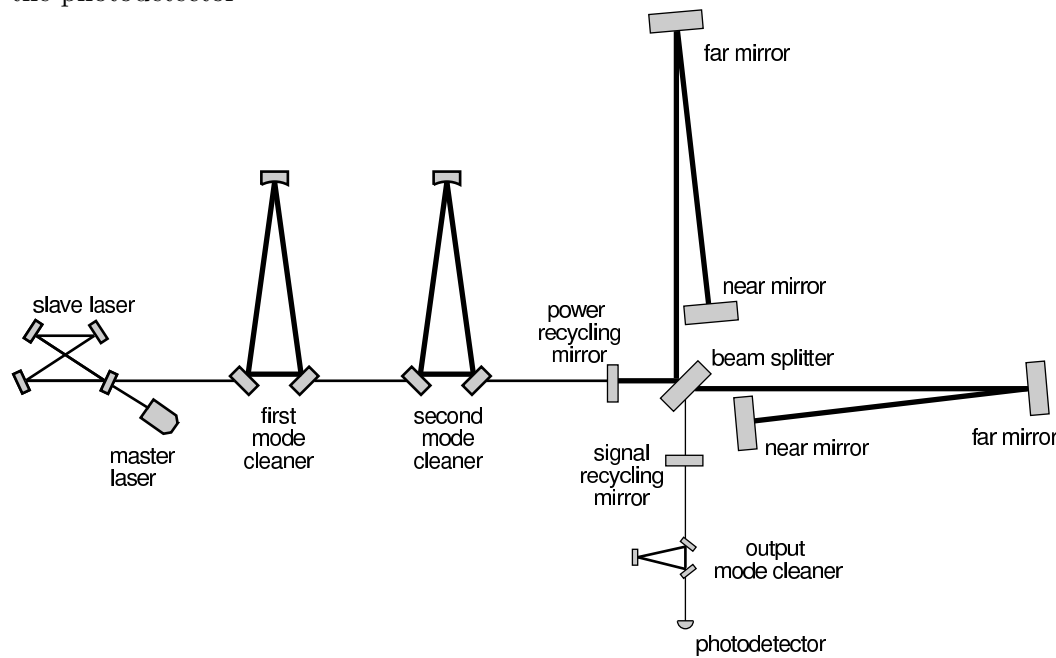}
\end{center}
\end{figure}

The  optical layout of GEO\,600 (see Figure
\ref{figure:GEO-layout}) can be divided into four  major parts:
The laser system, the input optics, the  dual-recycled Michelson
interferometer, and the output optics followed by the main
photodetector. Some steering mirrors, electro-optical modulators
and Faraday isolators are omitted in  Figure  7.  All optical
components but the laser system and the photodetector are
suspended inside the vacuum system.

\subsection{Laser}The  GEO\,600 laser system\cite{39} is based on an
injection-locked laser-diode pumped Nd:YAG system with an output
power of 12 W.  A non-planar ring-oscillator (NPRO) with an output
power of  0.8  W is used as the master laser. The slave laser is
formed by a four mirror  cavity with two Nd:YAG rods serving as
gain  media.  Each crystal is pumped by fiber-coupled laser diodes
with a power  of 17W. Two Brewster plates are incorporated in the
slave cavity  to define  the polarization direction, reduce
depolarization  losses and  compensate  for  the astigmatism
introduced  by  the  curved mirrors of the slave resonator. Due to
the mode-selective pumping scheme  more than 95\% of the light
leaving the laser  is  in  the fundamental  $\rm TEM_{00}$  mode.
The fully automated  injection  locking control servo system
acquires lock within 100 ms and allows stable operation of the
laser system.

\subsection{Input  Optics}The  light  from the  laser  system  is  passed
in transmission   sequentially  through  two   triangular resonant
cavities,  serving  as  frequency  references  and optical  mode
cleaners;  they are of 8.0 m and 8.1 m  round-trip lengths.   The
laser  frequency is stabilized to the resonant frequency  of  the
first  mode  cleaner MC1\cite{10}. For this purpose radio
frequency phase modulation  sidebands are impressed on the laser
beam prior to entering the first mode cleaner. The light reflected
by the input mirror  of  the  first  mode cleaner interferes  with
the light leaking  out  of  the first mode cleaner on a quadrant
photodiode. The  demodulated sum of  the  photocurrents of  all
quadrants  of this  photodiode is used in the Pound-Drever-Hall
scheme to develop an error signal for the deviation of the  laser
frequency from a mode cleaner resonance frequency. This signal is
fed  back to the master laser frequency actuators and to a  phase
correcting Pockels cell to stabilize the laser to the first  mode
cleaner length. With this first control loop in place, the  laser
frequency will change as the length of MC1 changes. Due  to  this
effect,  the length-control actuator of MC1 can be used to  bring
the laser/MC1 unit into resonance with the second  mode  cleaner
MC2. For this purpose another pair of rf sidebands is imposed  on
the laser beam by an electro-optical modulator,  located between
the two mode cleaners. The light reflected by the  input mirror of
the second mode cleaner is aligned onto a quadrant photodetector
and the sum of the photocurrents of all segments is demodulated to
produce an error signal for this feedback loop.  A third  control
loop is used to bring the laser/MC1/MC2 unit  into resonance with
the power-recycling cavity. A Faraday isolator  is used  between
the mode cleaner and the power recycling mirror  to obtain  access
to the light reflected by  the  power  recycling cavity  which  is
detected  on a quadrat  photodiode.  A  detailed description of
the frequency control scheme of GEO\,600  is  given in
Reference\,\cite{12}.

All  the  quadrant  photodiodes mentioned above are  used  both
for length  control  and  for wavefront  sensing  (and  thus
alignment control) of the mode cleaners and the power  recycling
cavity. The difference of the photocurrents of a combination  of
the quadrants of these diodes is demodulated at the respective rf
frequency and the resulting signals provide alignment information
of  the incoming beam relative to the eigenmode of the  relevant
cavity. A telescope  is used to get near field  and  far  field
information of the phase-front differences which can be  converted
into  tilt or rotation as well as $x$ or $y$ parallel shift
information of the  incoming  beam relative to the cavity axis
$z$. The appropriate linear  combination  of these signals is fed
back  to the  mode cleaner  mirrors.  The  alignment  error signal
of the power recycling cavity is used to change the tilt/rotation
of the power recycling  mirror  and of a steering mirror to keep
this cavity aligned  to  the incoming beam. The complete automatic
alignment system\cite{11}  uses additional quadrant diodes behind
several mirrors to  keep  the spot positions centered on all the
relevant cavity mirrors.

\subsection{Interferometer configuration}The main interferometer is
designed as  a  dual-recycled  folded-arm Michelson
interferometer\cite{7,8}.  Power recycling  leads  to  a power
buildup in the  interferometer  and improves the shot-noise
limited sensitivity of the detector.  The anticipated power
buildup in GEO\,600 is 2000 which results  in  a power  of about
10\,kW at the beamsplitter.   Any  differential phase change  of
the light in the  interferometer  arms  (the signature of a
gravitational wave) will lead to a change in the light intensity
at the  output port of the interferometer. The
partially-transmitting signal recycling mirror will reflect most
of this light back  into  the interferometer and  forms  another
Fabry-Perot cavity,  the signal-recycling cavity. In this cavity,
the  light power representing  the signal is  enhanced  through
resonance in  a frequency range determined by the cavity bandwidth
of  the signal recycling mirror  and the  resonant frequency  of
the signal-recycling cavity. This effect reduces  the
shot-noise-equivalent apparent displacement noise of the detector
for these frequencies. For the S1 run, the signal recycling mirror
was not yet installed, and so the instrument ran as a
power-recycled folded-arm Michelson.

In  the  final optical configuration GEO will use an output  mode
cleaner  as  a  spatial filter of the main interferometer  output
beam, placed just before the antisymmetric photodetector. The
output mode cleaner will be installed when the signal recycling
mirror is incorporated, but was not needed for the S1 run.

The  length and  alignment  control systems  for  the  Michelson
interferometer  and  the  signal  recycling  cavity  use  similar
techniques  as  described above for the mode cleaners  and  power
recycling  cavity.  Quadrant photodiodes  sense  the  beam  at
the interferometer  output port for the Michelson  control.  A
small fraction  of the light in one interferometer arm is
reflected  by the  anti-reflection coating of the beamsplitter and
is used  for the  control of the signal recycling degrees of
freedom. The  two pairs of sidebands needed for the sensing scheme
are impressed on the  laser beam injected into the power recycling
cavity, and the rf  frequencies were chosen to be multiples of the
free  spectral range of the power recycling cavity. Magnet and
coil actuators at the  intermediate  mass  of  the suspensions,
and  electro-static actuators  at the mirror level of the triple
pendulum suspensions, are used as actuators for the length and
alignment control loops.

GEO\,600  has five suspended cavities and the suspended Michelson
interferometer  which need length and alignment control  systems.
Thirty  pendulums need local damping of at least 4 degrees of
freedom and  additional feedback-control systems are needed for
the laser stabilization.  Most of these control loops are
implemented  with analog  electronic controllers with some
guidance  by  a  LabVIEW computer-control environment\cite{40}.
Only the active seismic isolation and some slow
alignment-drift-control systems are implemented  as digital
control loops.  The LabVIEW computer system controls
pre-alignment,  guides  lock acquisition of the laser  and  the
mode cleaners, monitors the detector status, and compensates for
long-term  drifts. Typical response times of this system are  100
ms. The   lock  acquisition  of  the  recycling cavities   and the
interferometer  is  guided  by a micro-controller  to  allow for
faster response times.

Although  only  the  light at the detector  antisymmetric  output
includes a possible gravitational-wave signal with a high  signal
to noise ratio, a multi-channel data acquisition system is needed
to  detect  environmental and detector disturbances  and  exclude
false detections.  Two different sampling rates (16384 Hz and 512
Hz) are used in the data-acquisition system (DAQ) of GEO\,600.  In
the  central  building 32 fast channels and 64 slow channels  are
available, and in each of the end buildings 16 fast channels  can
be  recorded.  Most of these channels will be used  for  detector
characterization  only. The data are recorded into  the  standard
Frame format for later analysis.

\section{The S1 Run} The  S1  run took place from 23 August 2002
15:00 UTC through  9 September 2002 15:00 UTC.  The total duration
of 17 days  spanned three weekends and one national holiday in the
U.S., which helped reduce the time lost due to anthropogenic noise
sources, particularly at  the  LIGO Livingston site.  Locked times
and duty cycles  for the  four individual interferometers are
given in Table \ref{table:lock-times},  along with  the  double
and triple coincidence times for  the  LIGO detectors.  The duty
cycle of GEO\,600 (98\%) is so high  that  its coincidence time
with any combination of LIGO interferometers  is essentially the
same as that of the LIGO interferometer(s) alone.

\begin{table}[!hbp]
\caption{Locked times and duty cycles for the S1 Run}
\label{table:lock-times}
\begin{tabular}{c|c|c}
\hline Detector/combination & Detector Hours coincidence  &
Locked/Duty
Cycle\\

LIGO H1          &          235    &    57.6\%\\

LIGO H2         &           298   &   73.1\%\\

LIGO L1         &          170   &        41.7\%\\

GEO\,600         &          400   &         98\%\\

H1-H2           &          188   &        46.1\%\\

H1-L1           &          116   &        28.4\%\\

H2-L1           &          131   &        32.1\%\\

H1-H2-L1        &           96   &        23.4\%\\

\end{tabular}
\end{table}

\subsection{LIGO}Each  LIGO  interferometer had a  defined  operating
state, including  which servos were operational, their gains,
acceptable light  levels  on photodiodes, etc.  The instrument
operator  on duty  was responsible to lock the interferometer and
put it  into the  required  configuration, assisted by computer
scripts that  set  the majority of the parameters.  When the
desired state was achieved, the  operator  issued a command that
put the interferometer  into ``Science  mode", effectively
declaring that the detector  was  in the   proper   configuration.
At  that time,  personnel   were restricted  from  entering the
experimental  halls.   A  computer program  began  monitoring the
computer control network  for  any unauthorized changes to the
interferometer state, and if any were detected,  it  automatically
removed  the  interferometer   from Science Mode and raised an
alarm.

Only  data taken in Science Mode segments longer than 300 seconds
were deemed suitable for analysis.  However, because of the still
incomplete  state  of the commissioning, the  designation  of  an
interferometer  as  being in Science Mode was not  sufficient  to
ensure  that  the data were of uniform quality.  Thus  each  data
analysis  effort  independently evaluated the Science  Mode  data
(assessing,  for example, the noise level or the quality  of  the
calibration)  and  made their own selection of data  for  further
analysis, based on the particular requirements of that analysis.

The  LIGO  noise level for S1 shown in Figure 8 is  substantially
above  the  design goal.  At high frequencies, most of the  extra
noise can be attributed to the fact that the interferometers were
effectively using very low laser power, and using the detection
sidebands inefficiently, as described above. This leads to a
poorer sensing resolution due to shot noise and electronics noise.
At  low  frequencies, the excess noise is mainly due to noise  in
the  control  systems, typically from auxiliary  control  systems
such  as  angular  controls. Much of this noise  is  due  to  the
incomplete  commissioning of the alignment  system,  and  control
system filters which are not yet optimized.

\begin{figure}
\caption{The LIGO interferometer sensitivities for S1. The LHO 4km
instrument is H1; the LHO 2km is H2; and LLO 4km is L1. The `SRD
Goal' refers to the LIGO design sensitivity for the LIGO
instruments.} \label{figure:LIGO-sensitivity}
\begin{center}
\includegraphics[angle=-90, width=1.0\textwidth]{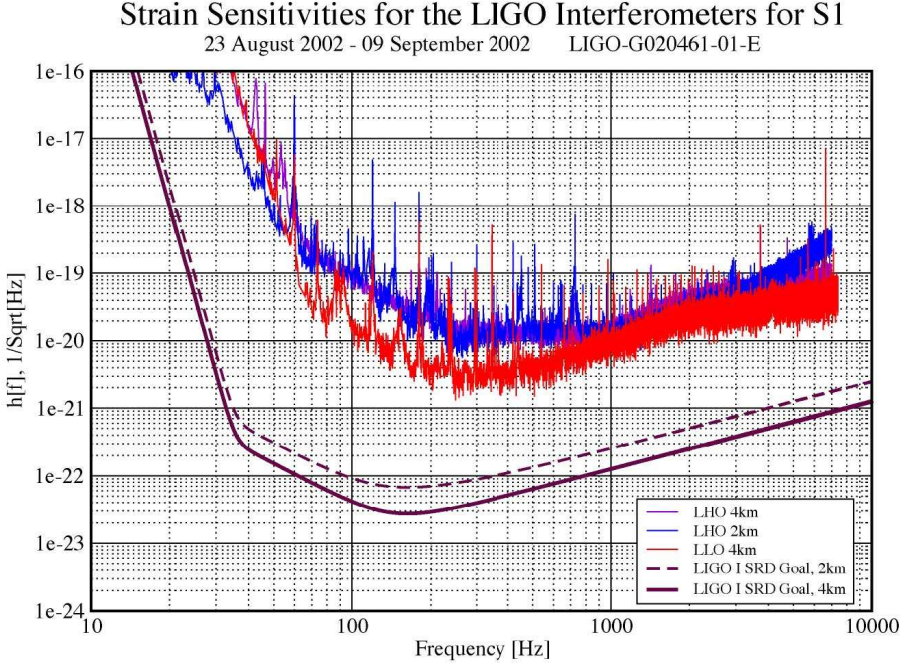}
\end{center}
\end{figure}

The  numerous  peaks  in the spectrum are  due  to  a  number  of
different sources.  Multiples of 60 Hz are prominent (at the
$10^{-19}$ level in strain), in part due to switching power
supplies that are scheduled for replacement.  Acoustic peaks, due
to fans and other rotating   mechanical  equipment,  enter through
acoustic and mechanical   coupling  mechanisms. Improvements in
acoustic shielding   and  in  the  optical layouts  to  reduce
acoustic sensitivity are planned to address these peaks.

The  signal  that  is  analyzed to search for gravitational  wave
signals  is  obtained  from the photodiode at  the  antisymmetric
port.  Since  this  is the error signal for the differential  arm
length, the effect of the feedback loop gain must be measured and
compensated  for.  The absolute scale for strain  is  established
using  the  laser wavelength, measuring the mirror  drive  signal
required  to move through a given number of interference fringes.
The  frequency response of the detector is determined via
swept-sine excitations of the end mirrors made periodically
through the run.

One  of the difficulties with the S1 data was that drifts in  the
alignment (because the wavefront sensing portion of the alignment
control system was not fully operational) caused changes  in  the
coupling  of  light  into  the  interferometer.   This  shows  up
directly  in  changes in the optical gain of the  interferometers
(e.g., watts per meter at the antisymmetric port), and in changes
of   the   overall   gain  of  the  servo  system   holding   the
interferometers in lock.  The first effect is an overall  scaling
in the signal, and was typically 10\% or less in S1.  However, the
change  in  servo gain gives a frequency-dependent correction  to
the  calibration function, which can be several times larger than
the  overall scale change, particularly near the unity gain point
of  the  differential arm servo (between 150  and  200  Hz).   To
compensate  for  this problem, the length of one of the arms was
modulated sinusoidally at two frequencies with known amplitudes
using the actuators on one of the end test masses. By   monitoring
the  size  of  the  resulting  signals  in   the antisymmetric
port photodiode signal, the optical  gain  of  the interferometer
could be tracked on a minute-by-minute basis,  and corrections for
the drift applied.  The calibration procedure and results are
described in more detail in reference [41]; the overall
statistical error is about $\pm$10\%.

Other measures of  performance also showed non-stationarity. The
band-limited rms  in   the antisymmetric port photodiode sometimes
showed degradation during a locked section  as the interferometer
drifted  away from  the initial alignment. ``Glitch" rates
included variations of factors of 3 to 100 between  locked
sections.   This non-stationary behavior affects  some searches
more than others, but should improve as the  detectors approach
full operation.

\subsection{GEO\,600}
\begin{figure}
\caption{The GEO\,600 Sensitivity Curve for S1.}
\label{figure:GEO-sensitivity}
\begin{center}
\includegraphics[angle=0,width=0.83\textwidth]{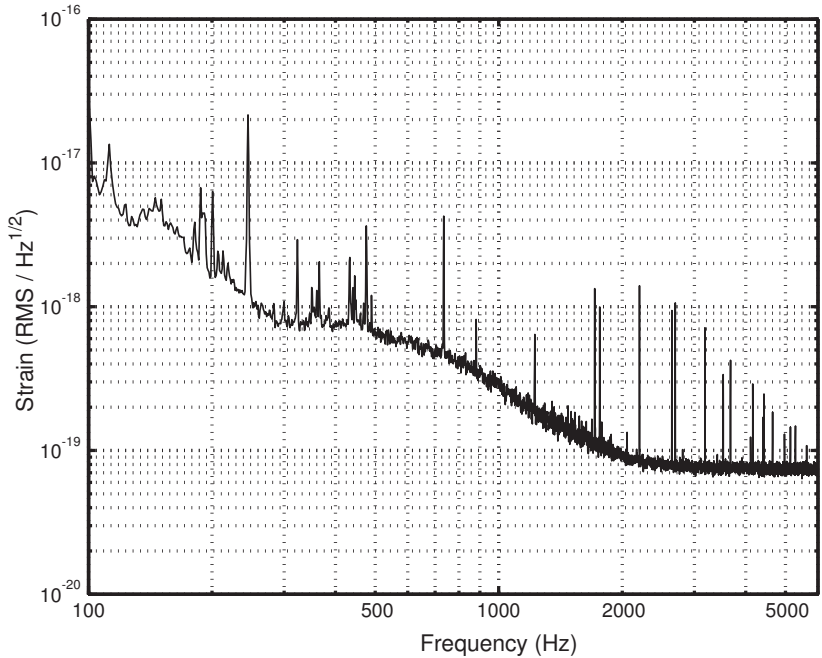}
\end{center}
\end{figure}

The  GEO\,600 sensitivity for the S1  run are shown in Figure
\ref{figure:GEO-sensitivity}. The duty cycle was 98\%  (see Table
3) and the longest continuous stretch of data is 121\,h. During
the S1 run GEO\,600 was operating in the power-recycled Michelson
interferometer configuration with  a reduced power recycling gain
and reduced input laser power.  The laser power injected into the
first mode cleaner was attenuated to 2W. The overall  optical
transmission of the  mode  cleaners, phase modulators and
isolators is 52\% which leaves approximately 1W  of laser  light
being injected into the interferometer.  The power buildup  in the
power recycling cavity was 300 which was limited by the 1.3\%
transmission of the power recycling mirror installed during  S1.
The signal recycling mirror was not installed  and test  mirrors
suspended in wire slings were used for the for  the beamsplitter
and the  inboard mirrors  of the  folded interferometer arms.

Due  to  the  automatic alignment system \cite{11}  the  longest
duration  without manual alignment of the mode cleaners was  more
than  one  year prior to S1 and no manual alignment of  the  mode
cleaners was undertaken during the S1 run. Operator alignment  of
the  power recycling cavity and the Michelson interferometer  was
needed  only a few times after major seismic disturbances.  An
automatic lock acquisition process was initiated whenever any of
the  cavities  or  the Michelson interferometer lost  lock.  This
system  and  the automatic alignment system were very stable  and
reliable and no operator presence at the detector was required at
night.

Calibration of GEO\,600 during the S1 run was achieved by imposing
known  forces  on  two of the mirrors using electrostatic  drives
that   were  fitted  to  provide  differential  control  of   the
interferometer.  In the final GEO\,600 configuration, radiation
pressure from a modulated laser beam which will be aligned onto
and reflected by one of the end mirrors will introduce calibration
lines with known amplitude into the output signal of the detector.
The spectrum of the applied calibration force consisted of a line
at 244 Hz and its odd harmonics, generated by suitable filtering
from a square wave. The signal generator used to produce the
series was phase-locked to  the  GPS stabilized clock to  which
the  data acquisition system  was  also  synchronized. The
electrostatic actuators were  calibrated  with  respect  to
electro-magnetic actuators one  stage  higher up on the same
suspension  chains. These were in turn calibrated by applying
forces large enough  to enable simple fringe counting.  The
actuators were found  to  be adequately  linear  to allow this
calibration method  to  succeed with  good reliability. The
calibrated gravitational wave channel was  generated  by measuring
the amplitudes of  the  calibration peaks  in  1 s frames of data.
These measurements were  used  to determine  the unknown
calibration  factors  (essentially   the overall optical transfer
function magnitude and the gain  of  the relevant control loops)
by fitting the data to a model based  on previously measured
transfer functions of the electro-mechanical control system. The
calibration coefficients were smoothed  over periods of one
minute.  Suitable time domain digital filters were generated to
produce a calibrated gravitational wave channel. The overall
calibration uncertainty was about 4\% for signal frequencies above
200 Hz and 6\% between 50 Hz and 200 Hz. Further detail  of this
process can be found in reference\cite{42}.

\section{Summary/future} The  S1  science run is an important
milestone for LIGO and  GEO, providing the first data for
scientific analysis for two  of  the newest  generation  of
gravitational wave interferometer observatories.   Even though the
detectors were operated in a preliminary configuration with many
features not implemented, the data were relatively well behaved,
and  the sensitivity great enough to improve  on  prior
observations with broadband gravitational wave detectors.

Several types of analysis have been recently completed using the
data from the S1 run. These include:

\begin{itemize}

\item    a search for the inspiral signal from binary neutron star
     mergers\cite{43}

\item  a search for continuous waves from a rapidly rotating
pulsar (J1939+2134)\cite{44}

\item   a search for short bursts of gravitational waves from
unknown sources\cite{45}

\item   a search for a stochastic background of gravitational
waves
     of cosmological origin\cite{46}
\end{itemize}

In  all  cases,  the sensitivity for S1 was not  expected  to  be
sufficient to make a positive detection, so the emphasis in these
analyses is to develop techniques for searching for gravitational
waves,  to  confront the problems of dealing with real data  with
its  deviations  from the usual assumptions  of  gaussianity  and
stationarity,  and to set improved upper limits on  the  flux  of
gravitational waves incident on the Earth.

Continued  rapid improvements are expected in both the  LIGO  and
GEO  detectors.  Immediately after the S1 run, GEO commenced  the
installation   and   commissioning  of   the   complete   optical
configuration  by  adding the signal recycling mirror.  LIGO  has
continued  to complete the control system and tuning, leading  to
more than a factor 10 improvement in the sensitivity. New science
data  runs  are  taking place to collect data  with  this  better
performance.  There will be a smooth transition from the  present
epoch,  where commissioning dominates, to the goal of effectively
continuous  astrophysical  data collection,  as  the  instruments
approach their goal sensitivity. The data are already interesting
in  terms of constraining astrophysical models from upper limits,
and  significant improvements in the near future will  make  even
better  upper limits possible, along with the increased potential
of   directly  detecting  gravitational  waves  of  astrophysical
origin.

\ack

The authors gratefully acknowledge the support of the United
States National Science Foundation for the construction and
operation of the LIGO Laboratory and the Particle Physics and
Astronomy Research Council of the United Kingdom, the
Max-Planck-Society and the State of Niedersachsen/Germany for
support of the construction and operation of the GEO600 detector.
The authors also gratefully acknowledge the support of their
research by these agencies and by the Australian Research Council,
the Natural Sciences and Engineering Research Council of Canada,
the Council of Scientific and Industrial Research of India, the
Department of Science and Technology of India, the Spanish
Ministerio de Ciencia y Tecnologia, the John Simon Guggenheim
Foundation, the David and Lucile Packard Foundation, the Research
Corporation, and the Alfred P. Sloan Foundation. This paper has
been assigned LIGO Laboratory document number LIGO--P030024.

\end{document}